\documentclass[aps,prx,reprint,superscriptaddress]{revtex4-2}
\usepackage{graphicx}
\usepackage{amsmath}
\usepackage{amssymb}

\begin{document}





\title{Magneto-synthesis effect on magnetic order, phonons, and magnons in single-crystal Sr$_2$IrO$_4$}



\author{Nicholas Pellatz}
\affiliation{Department of Physics, University of Colorado, Boulder, Colorado 80309, USA}


\author{Jungho Kim}
\affiliation{Advanced Photon Source, Argonne National Laboratory, Argonne, Illinois 60439, USA}

\author{Jong-Woo Kim}
\affiliation{Advanced Photon Source, Argonne National Laboratory, Argonne, Illinois 60439, USA}

\author{Itamar Kimchi}
\affiliation{School of Physics, Georgia Institute of Technology, Atlanta, Georgia 30332, USA}

\author{Gang Cao}
\affiliation{Department of Physics, University of Colorado, Boulder, Colorado 80309, USA}

\author{Dmitry Reznik}
\email[Corresponding Author: ]{dmitry.reznik@colorado.edu}
\affiliation{Department of Physics, University of Colorado, Boulder, Colorado 80309, USA}




\begin{abstract}
 

It was shown earlier that applying a magnetic field during the growth of Sr$_2$IrO$_4$, also known as "field-alteration", induces significant changes to its structural, magnetic, and transport properties. However, the microscopic nature of these changes is enigmatic. In this study, we employed resonant elastic and inelastic x-ray scattering, as well as Raman scattering, to investigate samples from two batches of Sr$_2$IrO$_4$ grown in magnetic fields of different strengths. Our findings reveal that samples grown in a weaker magnetic field have similar magnetic order to non-altered samples, whereas those grown in a stronger field show a different stacking of weak in-plane ferromagnetic moments. Additionally, we observed significant softening and broadening of select Raman-active phonons in the field altered samples, with a stronger effect in the samples grown in the stronger field. We discuss insights that our results provide into the microscopic nature of field-alteration in Sr$_2$IrO$_4$.

\end{abstract}



\maketitle

\section{Introduction}
Strongly spin-orbit-coupled, correlated materials offer an unparalleled playground in quantum materials research and engineering, due to the unique interplay between crystal structure and electronic charge, orbital, and spin degrees of freedom. A great deal of theoretical work addressing novel quantum states for these materials has thus far met very limited experimental confirmation. It has become increasingly clear that these discrepancies are due chiefly to the extraordinarily high sensitivity to their small crystal structural distortions/disorder. Here we studied a prototypical spin-orbit-coupled material structurally altered via application of magnetic field during crystal growth. The magnetic field aligns magnetic moments and, via strong spin-orbit interactions (SOI) and magnetoelastic coupling, crystal structures are altered as well at high temperatures. Like conventional crystal growth, the field altering technique generates crystals that are essentially consistent. This is demonstrated in our previous studies \cite{cao2020quest}. While still incomplete, results of an on-going investigation suggest the growth of a crystal is a strong function of the strength and orientation of the magnetic field.  

Application of magnetic fields can modify the Gibbs free energy such that phase boundaries become a function of the applied magnetic field, that is, $\Delta$G$_{Tot}$ = $\Delta$G$_{Chem}$ + $\Delta$G$_H$, where $\Delta$G$_{Chem}$ is Gibbs free energy in the absence of applied magnetic field H, and $\Delta$G$_H$ due to H. The magnetic energy is comparable with alloy mixing energies, ordering energies, and defect formation energies.  The magnetic field is particularly effective near a phase boundary where the Gibbs free energy difference between neighboring phases vanishes \cite{Rivoirard2013}. The total driving force for the transformation $\Delta$G$_{Tot}$ remains unchanged by the application of a magnetic field in the same cooling conditions, that is, $\Delta$G$_{Tot}$ (H=0) = $\Delta$G$_{Tot}$ (H>0). The relative contributions of $\Delta$G$_{Chem}$ and $\Delta$G$_H$ to $\Delta$G$_{Tot}$ (= $\Delta$G$_{Chem}$ + $\Delta$G$_H$) vary as a function of H. This relative change could be significant enough to induce a normally inaccessible phase, which has been recognized in previous studies \cite{Rivoirard2013}. For magnetic energy to be effective during a phase transformation, it is essential that the magnetic susceptibilities of the native and field-induced phases must be different. Over the last three decades, it has been established, both theoretically and experimentally, that magnetic energy can indeed change phase equilibria and stabilize metastable phases in functional materials, such as Fe-Co alloys, Nd$_2$Fe$_{14}$B, Nd$_2$Co$_{14}$B, SmCo alloys, etc. \cite{Rivoirard2013,guillon2018manipulation,gao2006effects}. This is particularly true for certain types of quantum materials with competing fundamental energies, such as Sr$_2$IrO$_4$.  

Iridium oxides or iridates have attracted a lot of attention in the recent decade due to potential to control their properties via the delicate interplay between strong SOI, electron-electron interactions and details of the crystal structure. 5d-electron orbitals of Ir are more extended in space compared to 3d, which increases hopping and, hence, the bandwidth. In addition, a larger orbital means that when there are two electrons in the same orbital, they are farther apart on average, which reduces on-site repulsion, $U$, the dominant electron-electron interaction in transition metal oxides. Based on these observations, iridates should be more metallic and less magnetic than materials based upon 3d and 4f elements, which have more compact orbitals \cite{jackeli2009mott}. The situation is much more complex when actual crystal structure and disorder are taken into account. For example, Ir-O-Ir bond angles in IrO$_{2}$ planes of the Ruddlesden-Popper phases, Sr$_{n+1}$Ir$_{n}$O$_{3n+1}$ (n = 1 and 2; n defines the number of Ir-O layers in a unit cell) are much smaller than 180$^{\circ}$, which strongly suppresses the hopping and reduces the bandwidth. Furthermore, strong SOI implies strong mixing between the charge and spin degrees of freedom that enhances the pinning of magnetic moments by the crystal lattice. The most profound manifestation of the novel physics of iridates is characterized by the $J_{\textrm{eff}} = 1/2$ Mott state in Sr$_{2}$IrO$_{4}$, which made it one of the most extensively studied materials. Due to obvious structural similarity with the parent of high temperature superconductors, La$_{2}$CuO$_{4}$, the Holy Grail of this extensive research effort is to develop a new platform for high temperature superconductivity. In fact, it is widely anticipated that Sr$_{2}$IrO$_{4}$ slightly doped with electrons should be a topological superconductor. When doped with Rh or La, Sr$_{2}$IrO$_{4}$ undergoes a metal-insulator transition, however, transport properties of the doped compounds are enigmatic and there were no hints of superconductivity \cite{qi2012spin,rau2016spin}.


An astonishing result is that an application of a small magnetic field of 0.8 T during crystal growth drastically alters transport and magnetic properties of Sr$_{2}$IrO$_{4}$ and its derivatives \cite{cao2020quest}. For example, (Sr$_{0.97}$La$_{0.03}$)$_2$IrO$_4$, which corresponds to 3\% electron-doping, remains  nonmetallic below 20 K for samples grown without an application of magnetic field, but, based on our still unpublished data, shows a rapid drop in electrical resistivity below 20 K for samples grown in the field.

The initial study has demonstrated that bulk properties of materials combining strong SOI with correlations grown under applied magnetic field are very different from the crystals of the same chemical composition grown without the magnetic field. Especially striking is the reduction of resistivity by several orders of magnitude in pure Sr$_{2}$IrO$_{4}$ and evidence for superconductivity in Sr$_{2}$IrO$_{4}$ lightly doped with La (G.Cao et al. unpublished).

Scattering experiments are crucial for characterizing magnetic degrees of freedom in these materials. We performed a systematic investigation of the effect of field altering on Sr$_{2}$IrO$_{4}$ by Raman scattering, resonant inelastic x-ray scattering (RIXS), and resonant x-ray scattering (RXS) to probe magnetic excitation spectra and phonons with high energy resolution. RXS uses elastic scattering to determine structure both atomic and magnetic.  RIXS is an inelastic scattering process much like Raman, but the higher momentum of x-rays compared to visible photons allows for measurements of excitations away from the zone center. We observed changes to certain magnetic excitations correlated with softening and broadening of certain Raman-active phonons as well as hardening of some other phonons. There was no measurable effect on apical oxygen modes, which indicates that structural modifications due to application of the magnetic field during crystal growth occur in the Ir-O planes.

\begin{figure*}
\centering
\includegraphics[trim={0cm 0cm 0cm 0cm},width=0.9\textwidth]{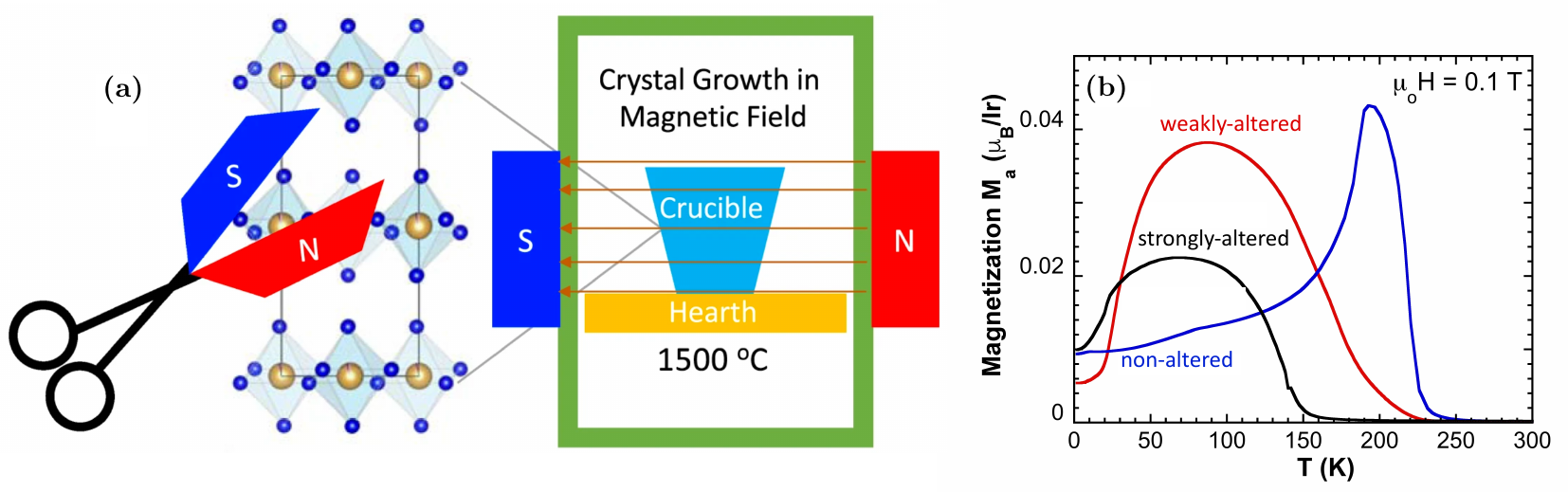} 
\caption{Cartoon of the field-alteration process (a), adapted from Ref. \cite{cao2020quest}.  Magnetization as a function of temperature for non-altered, weakly altered, and strongly altered samples (b).}
\label{mag-vs-temp}
\end{figure*} 

\section{Experimental Details}
Sr$_{2}$IrO$_{4}$ grown without applied field has recently been well-studied by several groups \cite{gretarsson2016two,gretarsson2017raman,PhysRevB.93.024405,PhysRevB.85.195148,PhysRevB.98.094101}. Field altered samples were grown using the flux technique under identical conditions as standard, except permanent magnets were placed on the outside of the growth chamber (Fig. \ref{mag-vs-temp}a) \cite{cao2020quest}. The field altered crystals are grown in a 1500 $^{\circ}$C-furnace carefully surrounded with either one or two permanent magnets, each of which is of 1.4 Tesla. The crystal samples grown with one magnet are “weakly” field altered and those with two magnets “strongly” field altered. The crystals were examined by energy-dispersive X-ray spectroscopy (EDX) and single-crystal x-ray diffraction to establish that field growth did not measurably affect chemical composition. Bulk magnetization (M) data suggest that $T_{N}$ is lowered to $\sim 210$ K in the weakly altered sample and $\sim 150$ K in the strongly altered sample (Fig. \ref{mag-vs-temp}b). $T_{N}$ was determined by the peak in dM/dT. There is no evidence that the magnetic field affects crystal orientation during growth.

Raman scattering experiments were performed with high quality single crystal samples mounted in a closed-cycle refrigerator. The spectra were measured with a custom-built McPherson triple spectrometer equipped with parabolic mirrors allowing measurements down to 5 cm$^{-1}$ in the crossed polarization geometry. Polarization of scattered light was analyzed with a polarizing cube, and all-mirror collecting optics were used.  For measurements of single-magnon and phonon scattering, we used a 532 nm laser and gratings with 1200 grooves/mm in the first and second monochromators and 1800 grooves/mm in the third monochromator.  For two-magnon scattering, we used a solid state 671 nm laser and lower-resolution gratings: 50 grooves/mm in the first and second monochromators and 150 grooves/mm in the third monochromator.  The notation $xx$ ($yx$) indicates that the incident and scattered light polarizations were aligned along the crystal axes and parallel (perpendicular) to each other.  In the $D_{4h}$ point group, the polarization geometries $xx$ and $yx$ measure excitations with symmetries $A_{1g}+B_{1g}$ and $B_{2g}+A_{2g}$, respectively.

Resonant x-ray scattering experiments were performed on beamline 6 at the Advanced Photon Source (APS) using standard beamline configuration \cite{chun2015direct}. Resonant inelastic x-ray scattering (RIXS) experiments were performed at beamline 27 at the APS. Incident x-ray energy of 11.2 keV that corresponds to Ir L$_3$ ($2p_{3/2}\rightarrow5d$) edge was used in the x-ray measurements.

\section{Results}
\begin{figure*}
\centering
\includegraphics[trim={0cm 0cm 0cm 0cm},width=1.0\textwidth]{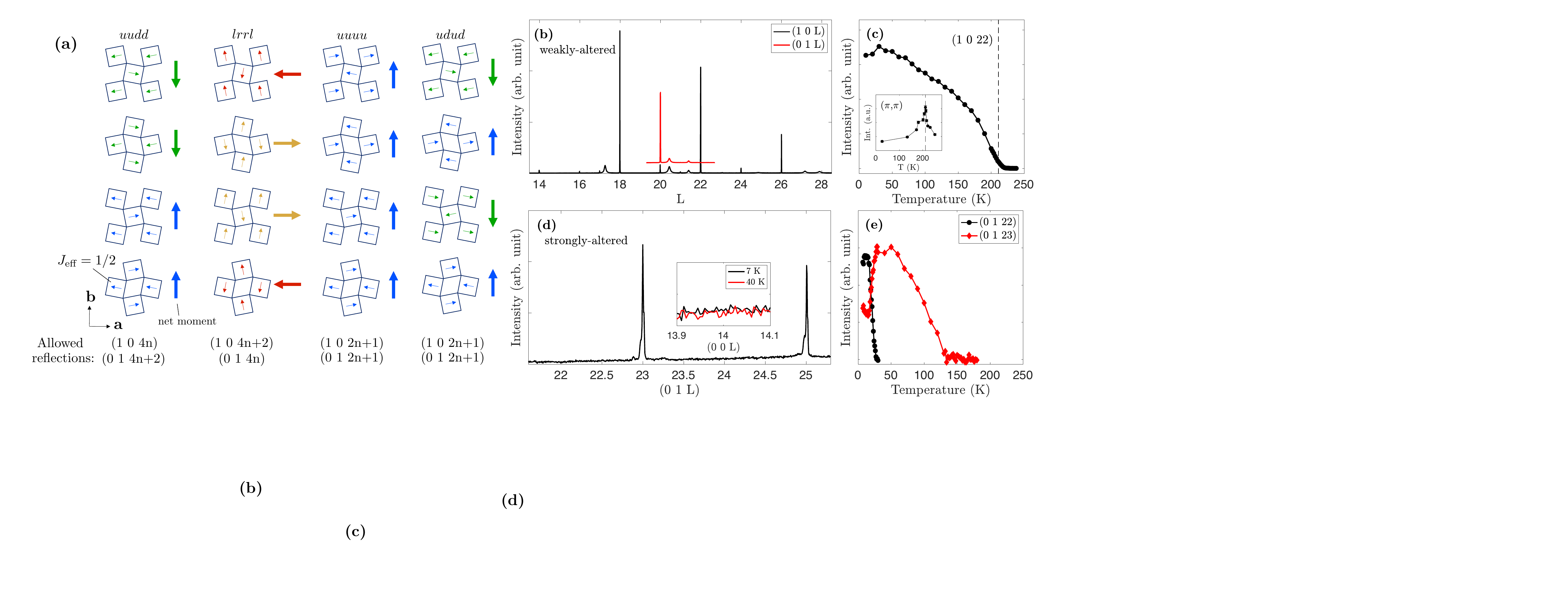} 
\caption{Top-down visualization of four possible stacking patterns of the $J_{\textrm{eff}}=1/2$ moments in Sr$_2$IrO$_4$ (a).  In each layer, the $J_{\textrm{eff}}=1/2$ moments add up to a net moment which points either up ($u$), down ($d$), left ($l$), or right ($r$) in the $ab$-plane.  Allowed reflections are listed below each stacking pattern.  Sr$_2$IrO$_4$ normally takes on either the $lrrl$ pattern or the $uudd$ pattern \cite{PhysRevB.99.085125}. (b) Typical RXS diffraction pattern in the magnetically ordered phase of the weakly altered sample. (c) The intensity of the (1 0 22) peak as a function of temperature (main panel) and magnetic fluctuations at the ($\pi$,$\pi$) point (inset). (d) Typical RXS diffraction pattern in the magnetically ordered phase of the strongly altered sample above 30K. This, coupled with the absence of a peak at (0 0 $2n$) (d, inset), points to the stacking pattern in the strongly altered sample being $uuuu$.  Panel (e) shows temperature dependence of two magnetic Bragg peaks in the strongly altered sample: (0 1 23), which is associated with $uuuu$ stacking and (0 1 22) which is associated with $uudd$ stacking.  The strongly altered sample appears to partially re-order to normal $uudd$ stacking below 30 K.\label{stacking-and-structure}}
\end{figure*}

Figure \ref{stacking-and-structure} demonstrates results of RXS measurements on field altered samples that elucidate the pattern of the magnetic order. Sr$_2$IrO$_4$ normally takes on either the $lrrl$ pattern or the $uudd$ pattern illustrated in panel (a) \cite{PhysRevB.99.085125}. The weakly altered sample shows magnetic Bragg peaks at (1 0 $4n + 2$) and (0 1 $4n$) in the magnetically  ordered phase, consistent with the normal $lrrl$ stacking observed in standard unaltered Sr$_2$IrO$_4$. The intensity of the (1 0 22) peak as a function of temperature (c) is proportional to the bulk magnetization in Fig. \ref{mag-vs-temp}b and the intensity of magnetic fluctuations at the magnetic zone center ($\pi$,$\pi$) (c, inset) show a sharp transition at 211 K. 

Above 30K magnetic Bragg peaks in the strongly altered sample (Fig. \ref{stacking-and-structure}d) are found at (0 1 $2n + 1$). This, coupled with the absence of a peak at (0 0 $2n$) (d, inset), points to the $uuuu$ stacking pattern in the strongly altered sample, which is different from standard Sr$_{2}$IrO$_4$. Below 30 K, this stacking coexists with the $uudd$ stacking observed as coexistence of the (0 1 23) associated with $uuuu$ stacking and (0 1 22) associated with $uudd$ stacking. So the strongly altered sample appears to partially reorder to normal $uudd$ stacking below 30 K. 

\begin{figure}
\centering
\includegraphics[trim={0cm 0cm 0cm 0cm},width=0.4\textwidth]{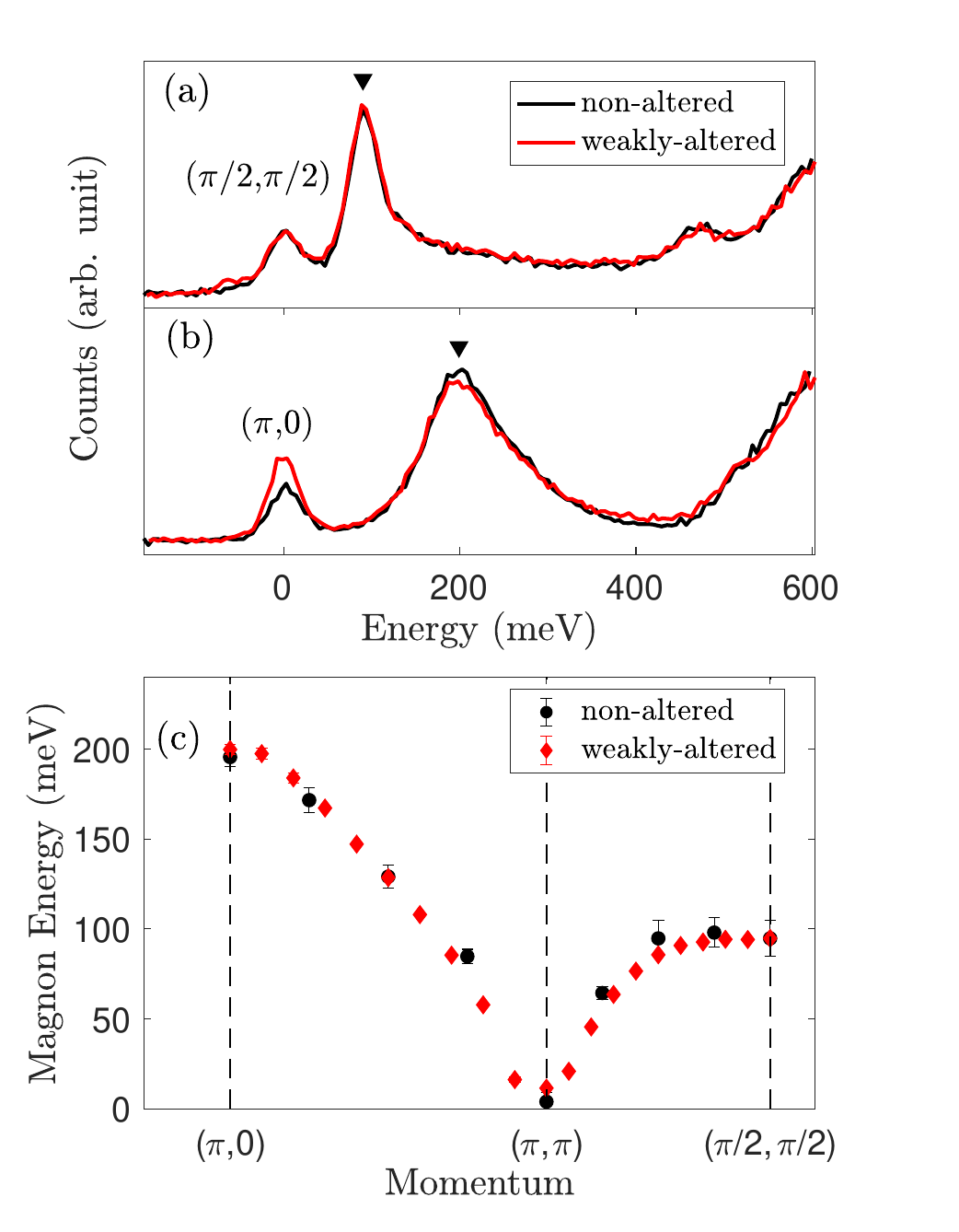} 
\caption{Raw RIXS data from a non-altered and weakly altered sample at ($\pi$/2,$\pi$/2) (a) and ($\pi$,0) (b).  Black triangles mark the positions of the peaks from one-magnon scattering.  At both zone-boundary points, the peak energies match nearly perfectly, suggesting that short-range in-plane couplings are unchanged in the weakly altered sample.  The magnon dispersion in (c) shows no major differences between the  weakly altered and non-altered samples.  Dispersion data on the non-altered sample are adapted from \cite{PhysRevB.101.094428}.}
\label{rixs_magnon_disp}
\end{figure}

\begin{figure}
\centering
\includegraphics[trim={0cm 0cm 0cm 0cm},width=0.4\textwidth]{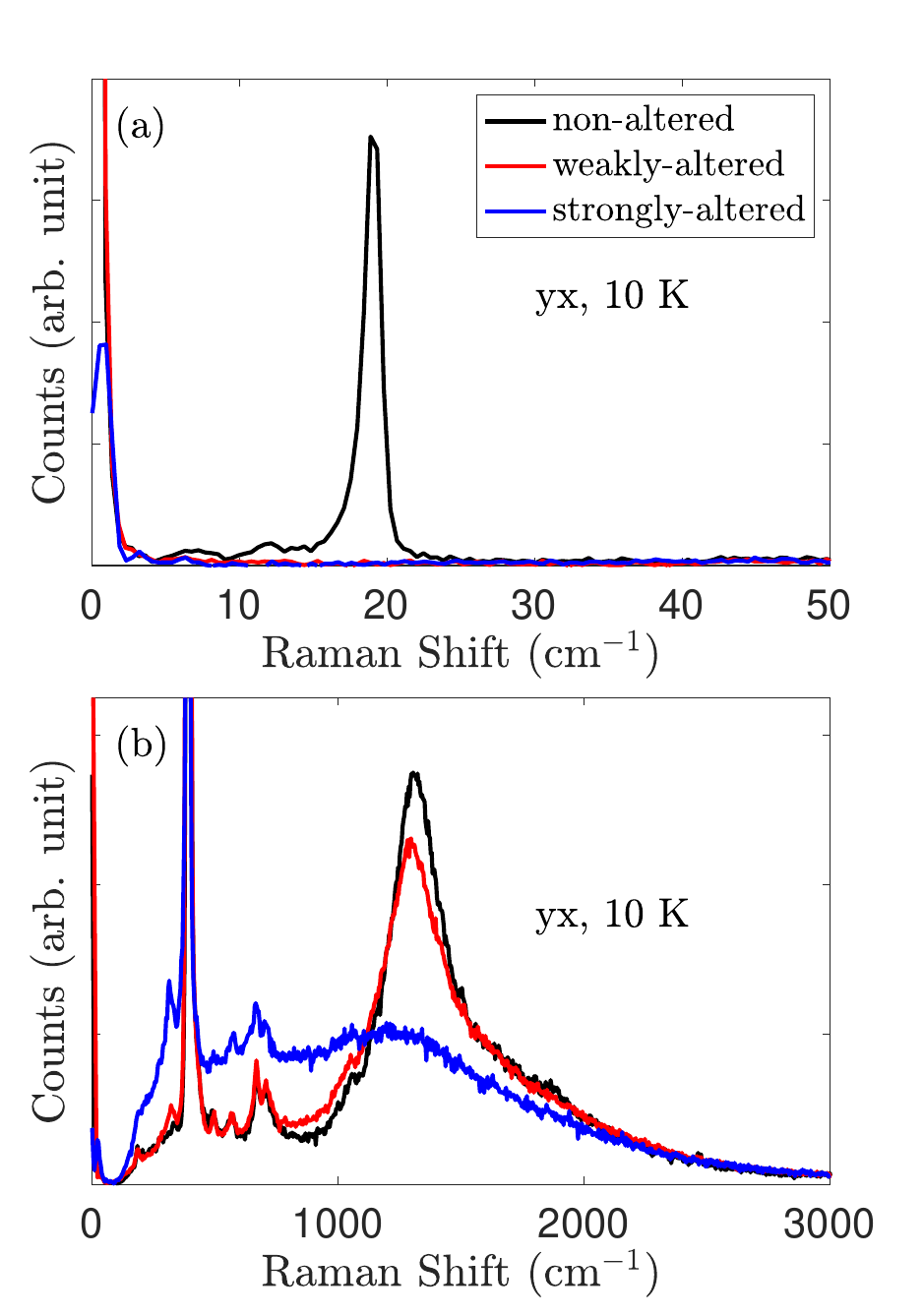} 
\caption{One-magnon (a) and two-magnon (b) Raman scattering at 10 K.  One-magnon scattering appears to be absent in the field altered samples, at least down to our energy cutoff of $\sim 5$ cm$^{-1}$.}
\label{raman_magnons}
\end{figure}

Magnon dispersions in the weakly altered sample were investigated by RIXS on beamline 27 at the APS, which is especially well-suited for measuring iridates. Figure \ref{rixs_magnon_disp} shows that the magnon spectra in the weakly altered samples are indistinguishable from those in standard Sr$_2$IrO$_4$. However, high resolution Raman scattering measurements demonstrate that the one-magnon peak clearly seen in standard Sr$_2$IrO$_4$ disappears from the spectrum of both weakly  and strongly altered samples (Fig. \ref{raman_magnons}a). In addition, the two-magnon Raman peak shows a small softening and broadening compared to unaltered Sr$_2$IrO$_4$. This effect is greatly amplified in the strongly altered sample whose two-magnon spectrum is considerably broader with the maximum appearing nearly 100 cm$^{-1}$ lower in energy (Fig. \ref{raman_magnons}b).

Since Raman scattering is a zone center probe, single magnons with nonzero wavevectors cannot be detected. Raman scattering is always inelastic so it can probe nonzero energies only. According to basic  spin wave theory, antiferromagnetic acoustic magnons should go to zero at the zone center making single magnons undetectable by Raman. In real materials, zone center magnon energies are nonzero due to crystalline anisotropy. This effect is magnified if spin-orbit coupling is strong, which is why the zone center magnon appears at a relatively high energy around 25 cm$^{-1}$ in the iridates at base temperature \cite{PhysRevB.93.024405}. In addition, spin-orbit coupling breaks inversion (anti) symmetry of the magnon excitation, which makes it Raman active. Based on this analysis, we conclude that disappearance of single magnons in the Raman spectrum can originate either due to the disappearance of magnetic order or due to the strong reduction of pinning of magnetic moments by crystalline anisotropy. The latter would send the zone center magnon energy to zero. Persistence of single magnons in field-modified samples close  to the zone center as seen by RIXS combined with magnetic order observed by RXS provides strong evidence for the latter mechanism.

Two magnon scattering originates mostly from magnons near the zone boundary. \cite{Fleury68,gretarsson2016two,gretarsson2017raman} Thus these are not the same magnons as those seen in the one-magnon Raman spectrum. The observation that strongly field altered sample has a profoundly modified two-magnon spectrum, whereas the weakly altered one does not is consistent with a different wavevector of magnetic order in the strongly altered sample, since a different order is expected to result in a different magnon dispersion. Investigation of magnon dispersion in the strongly altered sample will be performed in the future. 

In order to find out if structural changes occur as a result of field altering, we measured zone center phonons using Raman scattering in both strongly- and weakly altered samples. Raman scattering from phonons is sensitive not only to the average structure, but also to local structure such as oxygen vacancies. Ir and Sr appear at centers of inversion symmetry, so all Raman-active phonons are predominantly of oxygen character. Here we follow the phonon peak assignments in Ref. \cite{PhysRevB.98.094101}.

\begin{figure}
\centering
\includegraphics[trim={0cm 0cm 0cm 0cm},width=0.5\textwidth]{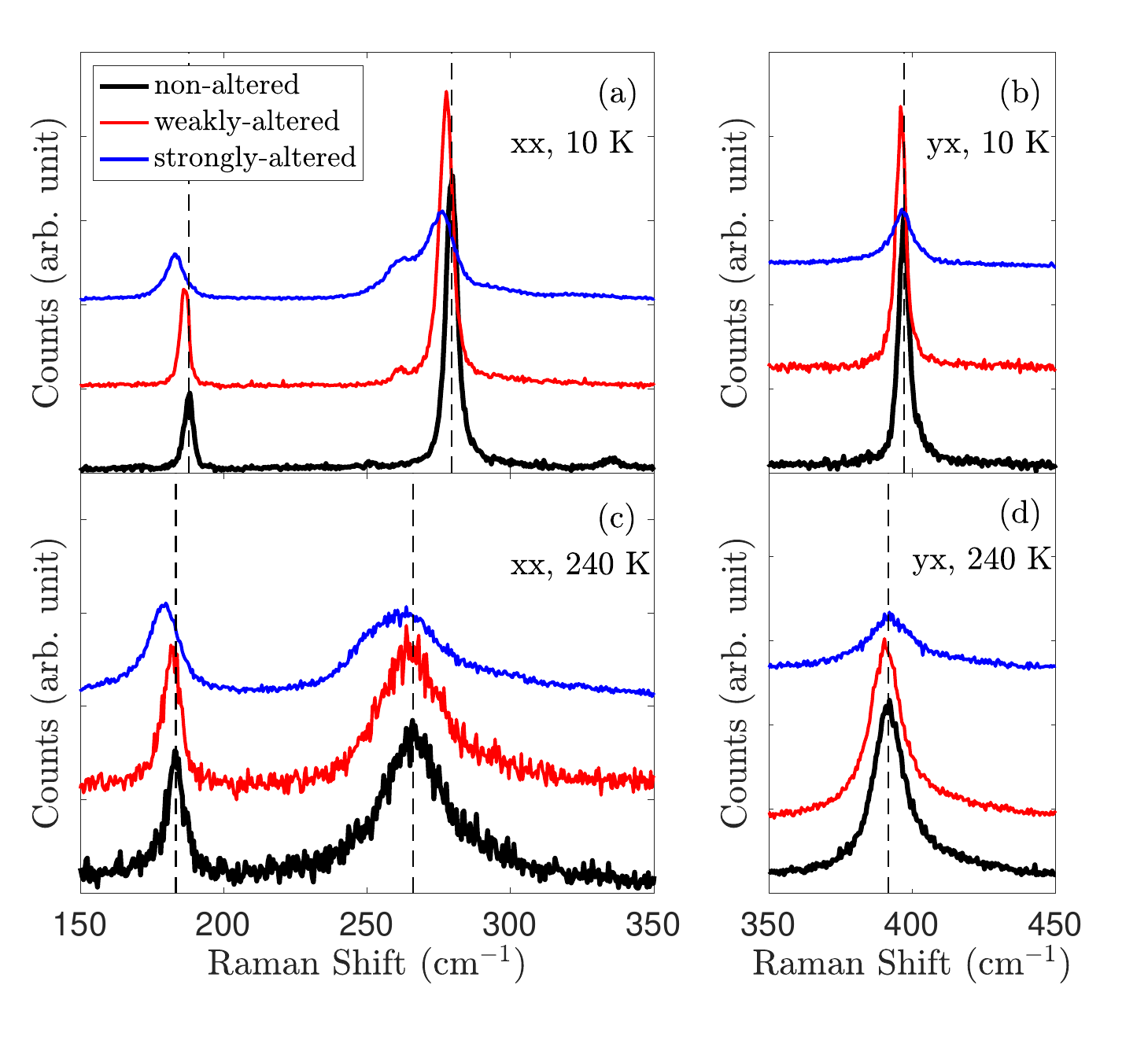} 
\caption{Raman scattering from phonons which bend the in-plane Ir-O bonds in the polarization geometries $xx$ ((a) and (c)) and $yx$ ((b) and (d)).  Vertical dashed lines indicate the energies of the phonons in the non-altered sample.}
\label{phonon_bending}
\end{figure}

We first examine phonons whose eigenvectors involve bending of the Ir-O-Ir bonds in the Ir-O planes. These include rotations of the Ir-O octahedra of $A_{1g}$ symmetry below 300 cm$^{-1}$ (Fig. \ref{phonon_bending} a,c) as well as so-called scissor modes of $B_{2g}$ symmetry where near neighbor oxygen atoms move against each other (Fig. \ref{phonon_bending}b,d). $A_{1g}$ phonons soften and broaden with field altering. This result is consistent with an earlier observation that Ir-O-Ir bond angle increases in field altered samples. Such an increase in the bond angle will result in smaller stretching of the bond lengths during the rotations, which will lead to smaller frequencies. In addition, the peaks broaden, especially in the strongly altered sample. A new mode is also visible at 10K at 260 cm$^{-1}$ in the field altered samples at low temperatures. It's intensity is enhanced  in the strongly altered one. This mode has been associated earlier with oxygen deficiency. We argue below that important differences exist between field altered samples compared with samples where oxygen deficiency was introduced by changing the growth conditions as opposed to just the application of the magnetic field \cite{sung2016crystal}. This mode is masked at 240 K where phonons broaden considerably in agreement with earlier studies. $B_{2g}$ ``scissor'' modes also associated with the bending of the Ir-O-Ir bonds appear at much higher energies (Fig. \ref{phonon_bending}b,c) and do not show consistent softening as a result of the application of the magnetic field during growth.

\begin{figure*}
\centering
\includegraphics[trim={0cm 0cm 0cm 0cm},width=0.7\textwidth]{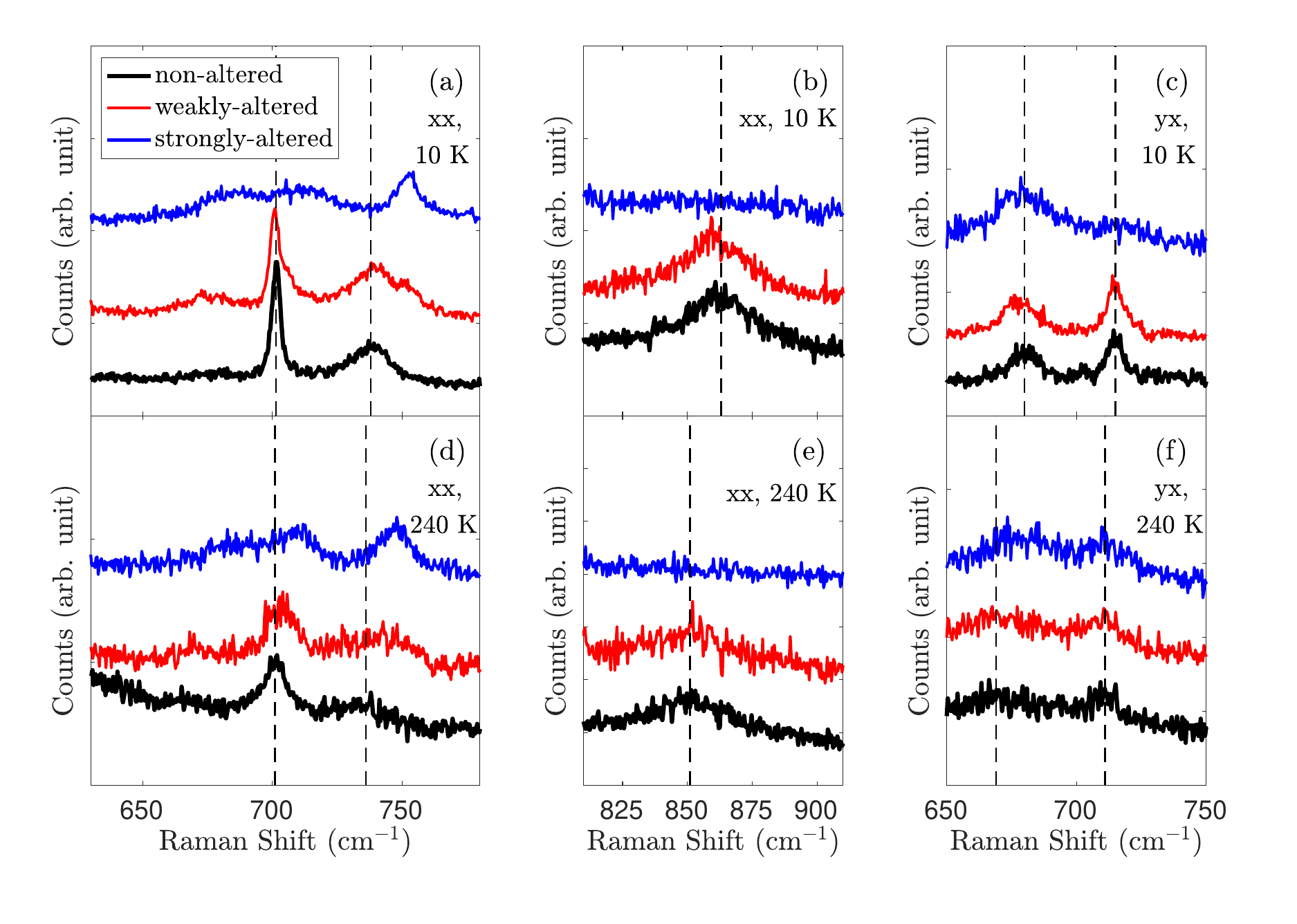} 
\caption{Raman scattering from phonon modes above 600 cm$^{-1}$ in energy.  We have not been able to assign an eigenvector to all of these modes, but based on their relatively high energy and the mode assignments in Ref. \cite{PhysRevB.98.094101}, we believe they must all involve stretching of the in-plane Ir-O bonds.  Vertical dashed lines indicate the energies of the phonons in the non-altered sample.}
\label{phonon_stretching}
\end{figure*}

Bond-stretching phonons appear at the highest energies around 700 cm$^{-1}$ (Fig. \ref{phonon_stretching}). Panel (a) shows that a new peak appears in the weakly altered sample around 750 cm$^{-1}$ in addition to the broad peak at 735 cm$^{-1}$. This peak disappears in the strongly altered sample and the new peak becomes prominent. The sharp peak at 700 cm$^{-1}$ acquires a shoulder on the high-energy side and a weak new peak appears at a lower energy close to 675 cm$^{-1}$. The main peak again disappears and only the new peaks remain in the strongly altered sample. Panel (b) illustrates that the phonon at 863 cm$^{-1}$ does not change between the weakly altered and unaltered samples, but disappears in the strongly altered one. The $B_{2g}$ peaks in panel (c) are largely unaffected by field altering.

\begin{figure}
\centering
\includegraphics[trim={0cm 0cm 0cm 0cm},width=0.5\textwidth]{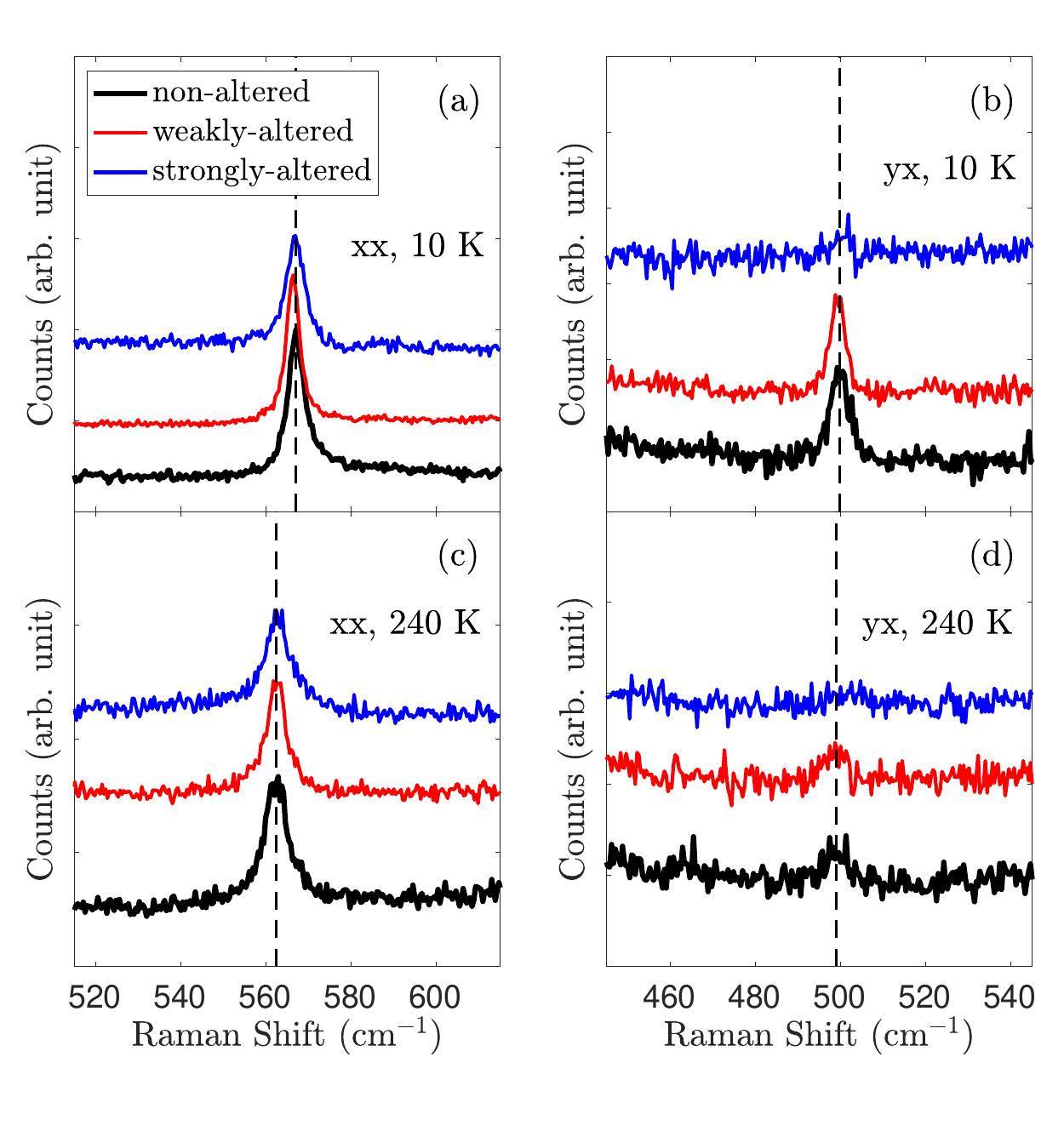} 
\caption{Raman scattering from the $A_{1g}$ (panels (a) and (c)) and $B_{2g}$ (panels (b) and (d)) apical oxygen modes.  These modes involve vertical movements of the apical oxygens and seem to be largely unaffected by field-alteration. }
\label{phonon_apical}
\end{figure}

Finally, Fig. \ref{phonon_apical} shows that the energies and linewidths of the apical oxygen modes are not affected by field altering as much as the modes discussed above. 

\section{Discussion}

\begin{figure}
\centering
\includegraphics[trim={0cm 0cm 0cm 0cm},width=0.4\textwidth]{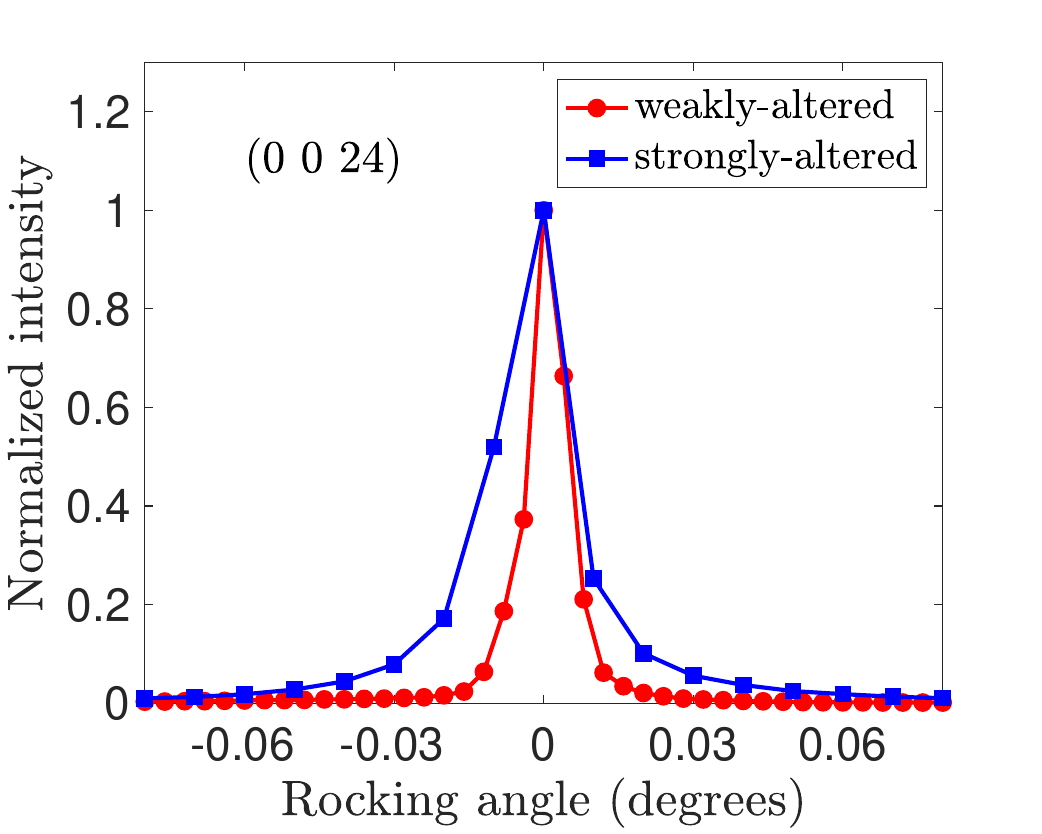} 
\caption{Rocking scans through the (0 0 24) structural Bragg reflection at 300 K.  The widths of these peaks are .007$^{\circ}$ and .014$^{\circ}$ for the weakly altered and strongly altered samples, respectively. The difference between the two samples is within systematic uncertainty of the measurement.}
\label{structural_bragg}
\end{figure}

Our results provide direct evidence that an application of the magnetic field during crystal growth strongly affects local structure in Sr$_2$IrO$_4$ in addition to inducing a relatively small change in the average Ir-O-Ir bond angle reported previously. For example two new phonon peaks appear in field altered crystals (at 260 and 750 cm$^{-1}$ in Fig. \ref{phonon_bending}a and Fig. \ref{phonon_stretching}a respectively). One possibility is that these modes are symmetry-forbidden in the unaltered crystals but become allowed in the field altered ones. However, IR reflectivity and ellipsometry measurements of IR-active but Raman forbidden modes \cite{propper2016optical} show that there are no phonon peaks at these energies. Therefore these are new modes resulting from local structural features different from local structure of stoichiometric Sr$_2$IrO$_4$. These modes are especially pronounced in strongly altered samples. At the same time structural Bragg peaks remain very narrow with rocking curve linewidths of under 0.05$^{\circ}$ (Fig. \ref{structural_bragg}), i.e. good crystallinity is retained.

Small effect of field altering on the apical oxygen phonons (Fig. \ref{phonon_apical}) provides another important clue: structural changes most likely do not occur in the Sr-O planes where apical oxygens sit, which leaves the Ir-O planes as the remaining possible hosts of structural modifications. Another clue is that the phonon at 260 cm$^{-1}$ has been associated with oxygen deficiency in an earlier study \cite{sung2016crystal}. These observations indicate that growing the samples in the magnetic field induces vacancies on the plane oxygen sites. Such vacancies will relax the structure reducing the octahedral rotations and inducing new vibrational modes in their vicinity. Increase in the rocking curve linewidth of the structural Bragg peak in the strongly altered sample also indicates increased oxygen deficiency. 

The field-synthesized samples are less distorted in terms of both the crystal and magnetic structures (which is the key advantage of magneto-synthesis), as such the Dzyaloshinskii-Moriya coupling, $D\cdot  (S_{i}\times S_{j})$, weakens (smaller S canting angels, lattice vector D, etc.), and the magnetic crystalline anisotropy reduces accordingly \cite{cao2020quest}. In addition, local disorder hinted at by the modifications of the phonon spectra should reduce the crystalline anisotropy responsible for the zone center one-magnon Raman peak, so its disappearance from the Raman spectrum is consistent with depinning of the magnons due to increased disorder. 
To clarify this point, this anisotropy serves as a potential well for magnetic moment directions in all unit cells. Classical equivalent of the zone center magnon excitation is a collective motion of these moments such that the directions of all moments in the crystal oscillate in phase inside this potential averaged over all unit cells. Note that in-phase nature of zone center modes eliminates any role of strong moment-moment interactions that determine magnon dispersions away from the zone center. Increase in the anisotropy will harden the frequency of the zone center vibration, whereas the decrease will soften it. If crystalline anisotropy is reduced to zero, the moments will be free to rotate together and the zone center magnon will be at zero energy. We argue that the data indicate that the anisotropy effectively disappears because Raman scattering cannot see excitations of zero energy.

Slight broadening and softening of the two-magnon peak in the weakly altered sample combined with a much larger effect in the strongly altered sample is reminiscent of the effect of oxygen deficiency induced by modified growth conditions reported earlier. The magnetic rearrangement transition of the strongly altered sample at 30K between uuuu and uudd (Fig. \ref{stacking-and-structure}e), is also seen in Fig. \ref{stacking-and-structure}c (weakly), and results in a loss of a net moment; it may be interpretable as caused by entropy favoring the thermal fluctuations of the net ordered moment at higher temperatures (a la order-by-disorder), but further experiments would be required to describe the rearrangement transition in detail.

It is not expected that a simple application of the magnetic field during growth can result in oxygen-deficient crystals, which are normally hard to make. This is because Ir ions tend to be tetravalent. A previous study indicates slightly oxygen-deificient Sr$_{2}$IrO$_{4}$ is possible only when the as-grown single crystals are annealed in vacuum at high temperatures for a long period of time. The single crystals would decompose when oxygen deficiency is more than 1\% \cite{qi2011electron}. 

Whereas we find similarities in the Raman spectra of our field altered samples with previously reported data on oxygen-deficient samples \cite{sung2016crystal}, there are important differences which lead us to believe that the striking properties of the field altered samples cannot be simply explained by in-plane oxygen vacancies.  First, rocking scans through a structural Bragg peak in oxygen-deficient samples found broadening to a width of about 0.03 degrees. This is significantly broader than the rocking curve of even the strongly altered sample (0.014$^{\circ}$), despite a similar prominence of the new 260 cm$^{-1}$ phonon mode. A different study found that the introduction of oxygen vacancies caused the unit cell volume to contract and the Ir-O-Ir bond angle to decrease with increasing temperature \cite{qi2011electron}. Both of these structural changes are opposite of what is observed in field altered Sr$_2$IrO$_4$ \cite{cao2020quest}. Finally, a metal-insulator transition was found in oxygen-deficient samples whereas there is no such transition in field altered samples. 

To summarize, spectroscopic and scattering techniques were used to investigate the microscopic changes in two batches of single crystals of Sr$_{2}$IrO$_{4}$ grown in different magnetic fields. It was found that applying a magnetic field during the growth changes properties of Sr$_{2}$IrO$_{4}$. The samples grown in a weaker magnetic field showed similar magnetic order to non-altered samples, whereas the samples grown in a stronger field had a different stacking of weak in-plane ferromagnetic moments. Raman-active phonons that correspond to the bending of the in-plane Ir-O-Ir bonds were significantly softened and broadened in the field altered samples, with a stronger effect in the samples grown in the stronger field. This suggests increased bond angles, which should increase the electronic bandwidth. There were some similarities between the Raman spectra of field altered and oxygen-deficient samples \cite{sung2016crystal}, but with increasing rather than decreasing Ir-O-Ir bond angles, and without a metal-insulator transition.

GC acknowledges NSF support via grant DMR 2204811. DR and NP acknowledge NSF support via grant DMR 2210126. This research used resources of the Advanced Photon Source, a U.S. Department of Energy (DOE) Office of Science user facility operated for the DOE Office of Science by Argonne National Laboratory under Contract No. DE-AC02-06CH11357.

\bibliography{fieldalter-214-refs}

\end{document}